\documentstyle[epsf]{aipproc}
\begin{document}
%****
\title{
SO(6)-Generalized Pseudogap 
%Pinned-Balseiro-Falicov 
Model of 
%Tunneling and Photoemission in 
the Cuprates}

\author{R.S. Markiewicz$^{1,2}$, C. Kusko$^{1,2,*}$, and M.T. Vaughn$^1$} 

\address{Physics Department (1) and Barnett Institute (2), 
Northeastern U.,
Boston MA 02115}
\maketitle
\begin{abstract}
The smooth evolution of the tunneling gap of Bi$_2$Sr$_2$CaCu$_2$O$_8$ 
with doping from a pseudogap state in the underdoped cuprates to a
superconducting state at optimal and overdoping reflects an underlying SO(6) 
instability structure of the $(\pi ,0)$ saddle points.  The pseudogap is
probably not associated with superconductivity, but is related to competing
nesting instabilities, which are responsible for the stripe phases.
\par
We earlier introduced a simple {\it Ansatz} of this competition in terms of a
pinned Balseiro-Falicov (pBF) model of competing charge density wave and 
(s-wave) superconductivity.  This model gives a good description of the phase 
diagram and the tunneling and photoemission spectra.  Here, we briefly review 
these results, and discuss some recent developments:  experimental
evidence for a non-superconducting component to the pseudogap; and SO(6)
generalizations of the pBF model, including flux phase and d-wave 
superconductivity. 

\end{abstract}

%\pacs{PACS numbers~:~~74.20.Mn, 74.72.-h, 71.45.Lr, 74.50.+r }

%****
%\narrowtext
%****

%\section{Introduction}

Recent photoemission\cite{Gp0,Gp2} and tunneling\cite{tu1,tu3} studies in 
underdoped cuprates find a remarkably smooth evolution of the pseudogap into the
superconducting gap as doping increases.  This has led to the suggestion that 
the pseudogap is caused by superconducting fluctuations or precursor 
pairing\cite{Rand}.  We suggest alternatively that the pseudogap represents a 
{\it competing} ordered state closely related to the stripe phases, with the 
smooth evolution due to an underlying SO(6) symmetry of the instabilities of 
the Van Hove singularity (VHS).
\par
In this picture, the stripe phases represent a nanoscale phase separation, 
between a magnetic (spin-density wave or flux phase) instability at half 
filling and a charge-density wave (CDW) near optimal doping\cite{Pstr}.
We have introduced a simple {\it Ansatz}, the pinned Balseiro-Falicov 
(pBF)\cite{MKK,BFal} model, which captures the essential features of the
stripe-superconductivity competition.  

%\section{Pseudogap Phase Diagram}

\begin{figure}
\leavevmode
   \epsfxsize=0.66\textwidth\epsfbox{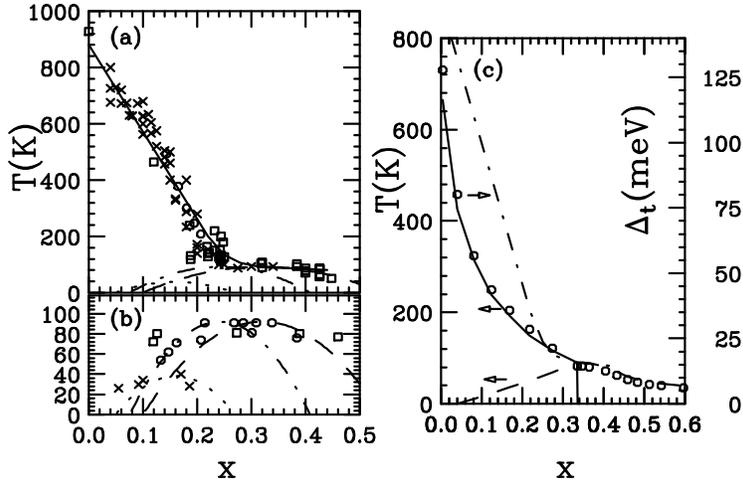}
\vskip0.5cm 
\caption{(a) Pseudogap phase diagram for Bi-2212 (squares) determined from 
photoemission\protect\cite{DCN}, and LSCO ($\times$'s) and YBCO (circles) 
determined from transport\protect\cite{BatT}.  (YBCO data
shifted by 10\% for better scaling.) Solid line = guide to the eyes; other lines
= parabolic approximations to superconducting $T_c$ for LSCO (dotted line),
YBCO (dashed line) and Bi-2212 (dot-dashed line).
(b) Uemura plot, with $n/m$ scaled to $x$.  Symbols and
curves have same meanings as in (a).
%}
%\begin{figure}
%\leavevmode
%   \epsfxsize=0.33\textwidth\epsfbox{tu70a.ps}
%\vskip0.5cm 
%\caption{
(c) Model pseudogap phase diagram for Bi-2212. Solid line = CDW transition
$T_p$; dashed line = superconducting transition $T_c$; circles = total gap 
$\Delta_t$ at 1K; dotdashed line = solid line from Fig.\protect\ref{fig:5a}a.}
\label{fig:5a}
\end{figure}

{\bf Pseudogap Phase Diagram:}
By comparing simultaneous measurements\cite{DCN} of the photoemission gap 
$\Delta$ with the pseudogap onset temperature $T^*$, we find an approximately 
constant ratio $2\Delta (0)/k_BT^*\simeq 8$, which allows us to plot the Bi-2212
pseudogap phase diagram as $T^*$ vs $x$, providing a direct comparison with
transport-derived pseudogaps in LSCO and YBCO\cite{BatT}, Fig.~\ref{fig:5a}a.
Remarkably, all three materials scale onto a single, universal phase diagram,
the scaling involving only a shift of the x-axes, relative to LSCO.  Such a
shift is, however, not consistent with a universal scaling of the 
superconducting $T_c$'s -- optimal T$_c$ falls at a different $x$ for each
material (parabolic curves in Fig.~\ref{fig:5a}a,b).  On the other hand, the 
Uemura plots\cite{Uem} also find that optimal $T_c$ falls at very different 
values of $n/m$ for different cuprates.  The simple assumption\cite{MG} that 
$n/m\propto x$ (with the constant of proportionality fixed by the LSCO data), 
{\it unifies the Uemura plot (symbols in Fig.~\ref{fig:5a}b) with the pseudogap 
scaling of $T_c$ (curves, Fig.~\ref{fig:5a}b)}.  This strongly suggests that the
scaling for YBCO and Bi-2212 merely converts the data to the correct value of x.
The resulting phase diagram can be well fit by the pBF model, 
Figure~\ref{fig:5a}c, although the ratio $2\Delta /k_BT^*$ is 4.1, somewhat 
lower than experiment.
%\label{fig:15}
%\end{figure}
%\par\noindent

%\section{SO(6)}

{\bf SO(6):}
The group structure of the model should be thought of not as a symmetry group,
but more in a renormalization group sense, as in the one-dimensional metal
g-ology.  (The group structure of g-ology has been discussed by Solomon and 
Birman\cite{SoBir}.)  Due to the logarithmic divergence of the density of states
near a VHS, the Fermi surface {\it almost} reduces to two
points -- the VHS's at $(\pi ,0)$ and
$(0,\pi )$.  The possible instabilities of the model have an underlying SO(6)
symmetry\cite{SO6}, but which instabilities are observed depend sensitively on
the form of the coupling constants -- corresponding to the g's of g-ology.
There are fundamentally two classes of instability -- nesting instabilities
which couple the two VHS's and pairing instabilities, which are intra-VHS.
\begin{figure}
\leavevmode
   \epsfxsize=0.66\textwidth\epsfbox{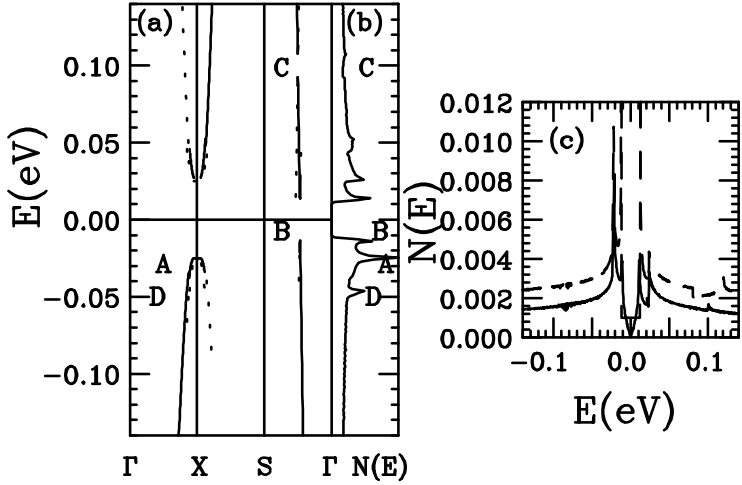}
\vskip0.5cm 
\caption{(a) Energy dispersion near Fermi level for underdoped 
cuprate\protect\cite{MKK}, illustrating spectral weight.  Coherence factor 
$\ge 0.6$: solid lines; between 0.1 and 0.6:
dashed lines; between 0.001 and 0.1: dotted lines. (b) Tunneling dos.
%\label{fig:3}
%\end{figure}
%\begin{figure}
%\leavevmode
%   \epsfxsize=0.33\textwidth\epsfbox{vck1.ps}
%\vskip0.5cm 
%\caption{
(c) Comparing tunneling dos for CDW-s-wave superconductivity (dashed line,
shifted up by 0.001) and flux phase -- d-wave superconductivity (solid line).}
\label{fig:3}
\end{figure}

The SO(6) symmetry of the model is most
clearly manifested in the equation for the total gap at $(\pi ,0)$: 
\begin{equation}
\Delta_t=\sqrt{\sum_i\Delta_i^2}, 
\label{eq:1c}
\end{equation}
where the $\Delta_i$ are the individual gaps associated with each instability.
(Note the g-ology flavor of this result: there is no underlying symmetry which 
says that all the $\Delta_i$'s are equal.)  In this case, the pBF model amounts
to the replacement $\Delta_{SDW}^2+\Delta_{CDW}^2\rightarrow\Delta_p^2$, where
$\Delta_p$ is the net pseudogap, which has similar form to a CDW gap.
Equation~\ref{eq:1c} shows that the smooth evolution of the tunneling gap with
doping is consistent with a crossover from magnetic behavior near half filling 
to superconducting behavior at optimal doping.

%\section{Photoemission and Tunneling Spectra}

{\bf Photoemission and Tunneling Spectra:}
Fig.~\ref{fig:3} compares the energy dispersion (a) and the tunneling spectra 
(b) near the Fermi level, in the underdoped regime.  
It can be seen that structure in the tunneling dos is directly related to
features in the dispersion of the gapped bands.  Thus, peak A is associated with
the dispersion at $(\pi ,0)$ -- the VHS peak split by the combined
CDW-superconducting gap.  Peak B is due to the superconducting gap away from
$(\pi ,0)$ -- particularly near $(\pi /2,\pi /2)$.  Feature C is associated 
with the CDW gap $G_k$ near $(\pi /2,\pi /2)$. 

\begin{figure}
\leavevmode
   \epsfxsize=0.33\textwidth\epsfbox{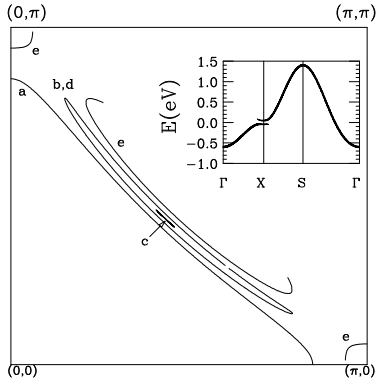}
\vskip0.5cm 
\caption{Constant energy surfaces for mixed flux phase -- d-wave 
superconductivity: energy = -50 (a), -20 (b), -1 (c), 20 (d), and 50 meV (e).
Inset: Corresopnding dispersion, with thickness of line proportional to 
coherence factor.}
\label{fig:3a}
\end{figure}

An equation similar to Eq.~\ref{eq:1c} arises in the theory of Bilbro and
McMillan\cite{BM} -- also a (three-dimensional) VHS theory -- and was postulated
to explain thermodynamic data on the pseudogap\cite{Lor}.  What is new here is
that the vector addition is found to hold only near the saddle points, while
the gaps split near $(\pi /2,\pi /2)$, and only the superconducting gap is near 
the Fermi surface there.  This is consistent with Panagopoulos and 
Xiang\cite{PX}, who found that, near the gap zero at $(\pi /2,\pi /2)$, the 
slope of the gap scales with $T_c$, and not with the gap near $(\pi ,0)$.  
Similarly, Mourachkine\cite{Mou} has found evidence for two tunneling gaps, 
very similar to features A and B of Fig.~\ref{fig:3}b; as predicted, feature B 
scales with T$_c$, and not with the pseudogap, feature A.  Feature B arises
from the superconducting gap at the hole pockets, as can be seen from the
energy dispersion at the B gap energy, curve b,d in Fig. \ref{fig:3a}.

The phase diagram is most naturally fit by assuming that the pseudogap
represents a nesting instability, and superconductivity a pairing instability.
A more precise determination will require careful experimentation.  Thus,
Fig.~\ref{fig:3}c compares the tunneling spectra for two models, the original
pBF model in terms of a CDW and an s-wave superconductor, and a modified
version involving flux phase -- d-wave superconductivity competition.  The 
resulting spectra are, as expected, nearly identical.  Close inspection shows
differences near $(\pi ,\pi )$, where the gap is purely due to the pairing
instability.  

%\section{Van Hove Pinning}

{\bf Van Hove Pinning:}
An essential ingrediant of the model is that the gap remains centered at $(\pi
,0)$ over the full doping range from half filling to optimal doping -- that is,
that the VHS is pinned at the Fermi level.  This remarkable consequence of 
strong correlation effects was first pointed out in 1989\cite{RM3}, and has
been rederived numerous times since then\cite{Surv1}.  We have noted that this
pinning should be measurable, both in tunneling and in photoemission, and a
preliminary analysis of the data appears to confirm the prediction\cite{MKK}.

%\section{Evidence for a Nonsuperconducting Pseudogap}

\begin{figure}
\leavevmode
   \epsfxsize=0.33\textwidth\epsfbox{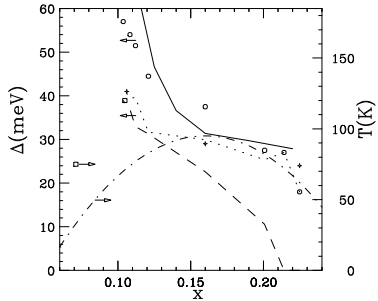}
\vskip0.5cm 
\caption{Tunneling gaps in BSCCO.
Circles = net tunneling gap, $\Delta$; dotted (dashed) line = estimate for
$\Delta_s$ ($\Delta_p$); dotdashed line = $T_c$; $+$'s = $10I_cR,$ where $I_cR$ 
is the average $I_cR$ product\protect\cite{tu7}; squares = estimate of T$_{c0}$
from Ref.~\protect\cite{COre}; solid line: from Ref.~\protect\cite{MKK}. }
\label{fig:12}
\end{figure}

{\bf Evidence for a Nonsuperconducting Pseudogap:}
The photoemission and tunneling spectra near optimal doping have a very
characteristic form below $T_c$.  There is a sharp quasiparticle peak at an
energy $\Delta$ above (or below) the Fermi level, with a pronounced dip near
2$\Delta$, followed by a broad hump at higher energies.  The dip is most
probably associated with reduced quasiparticle scattering within the 
superconducting state, which terminates when pairbreaking sets in at energies
above twice the superconducting gap, 2$\Delta_s$\cite{CoCo}.  Recently,
Miyakawa, et al.\cite{tu7} showed how the tunneling gap in Bi-2212 evolves with
doping, scaling a series of tunneling curves to the respective $\Delta$'s.  
These curves show significant deviations from scaling of the dip feature
with the tunneling gap, which suggest that in the underdoped regime,
$\Delta_s<\Delta$.  By assuming that the dip scales exactly with $\Delta_s$,
it is possible to extract the doping dependence of $\Delta_s$, and 
correspondingly of $\Delta_p$, the non-superconducting component of the 
gap\cite{MK2}, Fig.~\ref{fig:12}.

Shown also is recent Terahertz data from Corson, et al.\cite{COre}, who extract 
a bare superconducting transition temperature $T_{c0}>T_c$ from a 
Berezinski-Kosterlitz-Thouless analysis of superconducting fluctuations. 
(The effective $T_{c0}$ is found to be strongly frequency dependent; the 
squares in Fig.~\ref{fig:12} are an estimate based on the highest frequency 
data, 600GHz.) Note that $T_{c0}$ is considerably smaller than the pseudogap 
onset and has very different scaling, actually decreasing with increased 
underdoping.  Indeed, this $T_{c0}$ is consistent with the values of $\Delta_s$ 
estimated in Ref.~\cite{MK2}, with the same ratio of $\Delta$ to $T_c$ as found 
for the total gap in the overdoped regime (where the nonsuperconducting 
component is absent).  The data suggest a rather modest pair-breaking effect of 
the stripes, reducing the optimal $T_c$ from $\sim 125K$ to $95K$.  

%\section{Conclusions}

{\bf Conclusions:}
The simple pinned Van Hove {\it Ansatz} for the striped pseudogap phase in the 
cuprates provides a detailed explanation for the phase diagram and the 
experimental tunneling and photoemission spectra.  In particular: 
(1) The fact that the tunneling peaks are experimentally found to coincide with 
the $(\pi ,0)$ photoemission dispersion\cite {tu3} shows that the $(\pi ,0)$ 
dispersion has a gap -- that is, that the pseudogap is associated with VHS 
nesting\cite{Pstr}.  (2) The interpretation is self-consistent, in that the
experiments seem to find that the Fermi level is pinned near the VHS in the
underdoped regime\cite{MKK}.  (3) The tunneling gap has
a characteristic asymmetry which vanishes at optimal doping; this is evidence
that optimal doping is that point at which the Fermi level exactly coincides
with the VHS\cite{MKK}.  (4) While there are superconducting fluctuations in the
underdoped regime, a large fraction of the pseudogap has a non-superconducting 
origin.  (5) Portions of the tunneling spectra associated with the Fermi surface
near $(\pi /2,\pi /2)$ show distinct scaling with $T_c$, not $T^*$.
(6) The pseudogap phase diagrams for Bi-2212, LSCO, and YBCO appear to be 
universal and consistent with the Uemura plot, while optimal doping $x_c$ 
varies from compound to compound.  
\par

Our interest in the tunneling studies was sparked by conversations with A.M. 
Gabovich.  We would like to thank NATO for enabling him to visit us. 
MTV's work was supported by DOE Grant DE-FG02-85ER40233.  
Publication 757 of the Barnett Institute.

{\bf $*:$} On leave of absence from Inst. of Atomic Physics, Bucharest, 
Romania

\end{document}